# Charge Filling-Dependent Valley Polarization and its Dynamics Driven by Correlated Interlayer Exciton States in $WSe_2/WS_2$ Heterobilayer

*Zhiwei Li[1], Haonan Wang[1], Yubei Xiang[1], Wataru Idehara[1], Hailin Liu[1],*

*Shinichiro Matano[1], Kenji Watanabe[2], Takashi Taniguchi[3], and Kazunari Matsuda[1]**

[1]Institute of Advanced Energy, Kyoto University, Uji, Kyoto 611-0011, Japan

[2]Research Center for Electronic and Optical Materials, National Institute for Materials Science, 1-1 Namiki, Tsukuba, Ibaraki 305-0044, Japan

[3]Research Center for Materials Nanoarchitectonics, National Institute for Materials Science, 1-1 Namiki, Tsukuba, Ibaraki 305-0044, Japan

Corresponding author: Kazunari Matsuda

E-mail: matsuda@iae.kyoto-u.ac.jp

## Abstract

Moiré heterostructures of transition metal dichalcogenides provide an emerging platform for studying the strongly correlated electronic and excitonic states. The valley degree of freedom and its dynamics in the strongly correlated electronic and excitonic states in the moiré system are crucial for fundamental understanding and applications of valleytronics. Although the strongly correlated states in moiré systems have been extensively investigated, their impact on valley degrees of freedom and its dynamics remains unexplored. Here, we study the charge filling-dependent valley relaxation dynamics of interlayer excitons (IXs) in $WSe_2/WS_2$ moiré heterobilayer system. The evolution of moiré IXs dynamics across the Mott transition is elucidated by combining time-resolved optical spectroscopy with Monte Carlo simulation. Our results show that the radiative decay and recombination rate of IXs are enhanced at the integer filling factors $n\,(=-1\,)$. Moreover, the valley relaxation rate of IXs at $n=\pm1$ is significantly decreased. The resulting reduction in wavefunction overlap and phonon scattering process suppresses the nonradiative recombination and intervalley scattering of IXs. These anomalous behaviors are attributed to the modification of IXs recombination pathways and enhance excitonic coherence due to the localization of introduced charge carriers at integer filling of moiré site. Our results explore exciton and valley dynamics as sensitive probes of strongly correlated states in moiré systems and highlight possible ways to control excitonic coherence and valleytronic properties in two-dimensional quantum materials.

## Introduction

Monolayer semiconductors of transition metal dichalcogenides (TMDs) ($MX_2$: M=Mo, W; X=S, Se) have attracted considerable attention in fundamental physics and optoelectronic applications[1–5]. In monolayer TMDs, the inequivalent $+K$ and $-K$ valleys exhibit valley-contrasting optical selection rules under circularly polarized excitation[6]. Consequently, the corresponding valley pseudospin can be initialized and probed optically, making its control fundamentally important for valleytronic applications. Stacking monolayer transition metal dichalcogenides (TMDs) with a small twist angle generates a moiré superlattice and periodic potential[7–10], which can confine optically generated electron-hole pairs (excitons) in moiré sites[11–16]. The resulting moiré superlattice provides a versatile platform for studying the valley degree of freedom[17,18], because these valley properties are naturally inherited by the moiré excitonic states[19–22]. In particular, the valley relaxation dynamics of interlayer excitons (IXs) are crucial for valleytronic applications because they determine the duration for which circular polarization can be preserved against intervalley depolarization[23].

Recently, strong electronic and excitonic correlations originating from flat moiré minibands[24–26] and reduced dielectric screening[27–31] have been focused on the moiré system because the flattening of moiré minibands suppresses the electronic bandwidth and effective hopping between neighboring moiré sites, thereby enhancing the relative importance of Coulomb interactions[32–34]. By introducing charge carriers into the moiré site, the correlated electronic phases such as Mott insulating states at integer filling and Wigner crystal states at fractional filling have been experimentally observed in $WSe_2/WS_2$ heterobilayer[35,36]. Excitons in such moiré systems are highly sensitive to charge filling and dielectric screening, making them effective probes of correlated states. Previous studies have primarily relied on steady-state photoluminescence (PL) and reflectance spectroscopy to identify the signatures of correlated phases such as Mott transition in moiré systems[27,35], which only provide the limited static information and cannot directly resolve the underlying dynamics[37,38]. Thus, insights obtained from time- and polarization-resolved optical spectroscopy are essential for disentangling the intrinsic mechanisms of valley relaxation of moiré IXs and tracking their dependence on the charge filling of moiré sites[39], thereby offering a pathway toward understanding and controlling valley dynamics in strongly correlated states of moiré IXs systems.

In this study, we investigated the charge filling-dependent valley relaxation dynamics of IXs in the $WSe_2/WS_2$ moiré heterobilayer system. We elucidated how moiré IXs dynamics evolve across the Mott transition by combining time-resolved optical spectroscopy with Monte Carlo simulations. Our results show that the radiative decay and recombination rates of IXs are enhanced at the integer filling factor $n\ (=-1)$. Moreover, the valley relaxation rate of IXs at $n = \pm 1$ is decreased. These anomalous behaviors are attributed to modification of IXs recombination pathways and enhance excitonic coherence due to the localization of introduced charge carriers at integer filling of moiré site. These findings highlight an unconventional route for manipulating valleytronic properties by exploiting the strongly correlated states in moiré systems[40].

## Results and Discussion

**Figures 1(a)** and **(b)** show a schematic illustration and an optical image of the H1 device, respectively. Two H-stacking (60° twisted) $WSe_2/WS_2$ moiré heterobilayer devices (H1 and H2) encapsulated by hexagonal boron nitride ($h$-BN) to ensure a uniform and stable dielectric environment were fabricated using an all-dry transfer technique (shown in **the Methods section**)[11]. Few-layer graphene electrodes were employed to control the electrostatic doping process via dual-gate tuning. Throughout the experiments, the equal voltages were applied to the top and bottom gates ($V_g = V_{tg} = V_{bg}$) to tune the carrier density (filling factors) while maintaining a zero vertical electric field, thereby eliminating the influence of Stark shifts. All optical measurements were performed at 5.8 K.

**Figure 1(c)** presents the two-dimensional photoluminescence (PL) map of the $WSe_2/WS_2$ heterobilayer device as a function of gate voltage ( $V_g$ ), with the corresponding PL spectra provided in **Fig. S(1)**. The PL peak around 1.42 eV at $n \sim 0$ is observed, which originates from the emission of IXs in the H-stacked $WSe_2/WS_2$ heterobilayer. The PL intensity and peak position vary significantly with gate voltage ($V_g$). Pronounced changes in peak energy and intensity are clearly observed at $V_g$=1.5, and -3 V. With consideration of thickness of the $h$-BN dielectric layer, the electrostatic carrier (electron or hole) density is evaluated from the applied vertical gate voltage of $V_g$. Moreover, the filling factor $n$ defined by the number of electrostatically doped charge carrier density in each moiré potential is evaluated from $V_g$. The significant blueshifts of PL peak energy of 16.0 and 24.3 meV are observed at the critical filling of $n = -1$ and 1, respectively, as shown in the right axis of **Fig. 1(b)** in the PL map, providing unambiguous evidence of carrier-IX interactions associated with the formation of Mott insulator state[27]. A similar phenomenon was observed for device H2, as shown in **Supplementary Note 1**.

The PL peak energies extracted from the contour plot are depicted in **Fig. 1(d)**, obtained by fitting the PL spectra with a Voigt function. To understand these blueshifts, we use the Hubbard model to describe the electronic correlations in the $WSe_2/WS_2$ moiré system[29]. The formation of flat moiré minibands strongly suppresses the kinetic energy of carriers, thereby enhancing the importance of Coulomb interactions[24]. Electrostatically doped holes first occupy the valence band maximum (VBM), and the VBM splits into a lower Hubbard band (LHB) and an upper Hubbard band (UHB), separated by the Mott charge gap, owing to the strong correlation effects in the flat band. For filling factors $-1 < n \leq 0$, the electrostatic doped hole cannot fully occupy the lowest energy band LHB, so the hole of moiré IX can also reside in the LHB without additional energy cost. However, at the critical integer filling $n = -1$, where every moiré unit cell is already occupied by a hole and LHB is fully occupied, placing the hole of moiré IX into VBM (UHB) requires overcoming the energy separation of the Mott charge gap, which manifests as an abrupt blueshift in the PL spectrum. For electron doping, the situation is analogous, except that carriers first occupy the lowest conduction band (CBM).

We also addressed the electronic correlations of moiré IX and whether the excitonic coherence was modified by analyzing the linewidths of IX peaks[41,42]. As summarized in **Fig. 1(e)**, the spectral linewidths extracted from the analysis exhibit a pronounced narrowing at $n = -1$, indicating reduced homogeneous and inhomogeneous linewidth broadening arising from elongation of excitonic coherence and reduction of charge environment inhomogeneities. The abrupt changes in both spectral peak energy and linewidth in the PL spectra strongly suggest the excitonic Mott transition at the critical filling factor of $n = -1$.

**Figure 2(a)** shows a contour plot of the time-resolved PL decay curves in the 80 ns time window with different filling factors, revealing a clear evolution of the moiré IX lifetime. The PL decay exhibited clear biexponential behavior, consisting of a fast component on a few nanoseconds timescale and a slow component extending over several tens of nanoseconds. These decay curves were fitted using the following double-exponential function[43]:

$$I(t) = A_{\mathrm{f}}\exp\left(-\frac{t}{\tau_{\mathrm{f}}}\right) + A_{s}\exp\left(-\frac{t}{\tau_{\mathrm{s}}}\right), \tag{1}$$

where $A_{\mathrm{f}}\,(A_{\mathrm{s}})$ and $\tau_{\mathrm{f}}\,(\tau_{\mathrm{s}})$ denote the amplitude and lifetime of the fast (slow) components, respectively. The fitted amplitudes and decay rates are presented in **Fig. S3(a)** and **Fig. S3(b)**. Notably, at the strongly correlated state of $n = \pm 1$, both the fitted amplitude and decay time exhibit similar change as a function of filling factor. **Figure 2(b)** shows the effective lifetime $\tau_{\mathrm{eff}}$ of moiré IXs calculated by the equation of $\tau_{eff} = \frac{A_f\tau_f + A_s\tau_s}{A_f + A_s}$. In the electron-filling regime $(n > 0)$, the effective lifetime of IXs slightly decreases, whereas in the hole-filling regime $(n < 0)$, it exhibits an initial increase followed by a sharp decrease at the integer filling factor of $n = -1$.

The charge filling-dependent evolution of the IXs lifetime, together with the two decay components discussed above, suggests that the IXs dynamics involve multiple recombination pathways. To qualitatively describe these two recombination channels, we therefore introduce a three-level model, as illustrated in **Figure 2(c)**. The $WSe_2/WS_2$ heterobilayer has a type-II band alignment, with the VBM located in the $WSe_2$ layer and the CBM in the $WS_2$ layer. The previous studies have shown that, in the lowest conduction band of $WS_2$, the $Q$-valley has lower energy than the $K$-valley[2]. Upon optical excitation, electrons and holes are initially generated from ground state at the $K$-valley in VBM of $WSe_2$, and $K$-valley of CBM in $WS_2$, forms the IXs. Because the $Q$-valley lies in lower energy, the IX can have two distinct recombination channels. They can either recombine radiatively to the ground state or undergo intervalley scattering to the lower-energy $Q$-valley in the CBM of $WS_2$, where spin-flip relaxation assisted by phonon absorption occurs, followed by non-radiative decay to the ground state. Therefore, in the three-level model, the $K$- and $Q$-valley are regarded as bright and dark excitonic states, since the former is momentum-direct and the latter is momentum-indirect transition[44,45].

To go beyond the lifetime analysis and quantitatively determine the recombination and intervalley scattering rates, we developed a Monte Carlo model based on the Gillespie algorithm[43]. The photoexcitation is treated as a stochastic process following Poisson statistics in the model. Upon each excitation event, an IX is initialized in either the $K$- or $Q$-valley excitonic state with probabilities $p_K$ and $p_Q$, respectively. Since the model only considers the two excited states after photoexcitation, their probabilities are normalized to $p_K + p_Q = 1$. Accordingly, the initial population satisfies the relation of $N_K(0){:}N_Q(0) = p_K{:}p_Q$, while $N_0(0) = 0$. The weak excitation conditions, initial state assumptions, and temporal settings employed in the Monte Carlo simulation are detailed in **Supplementary Note 2**. The IXs dynamics are governed by the recombination rates $\Gamma_K$, and $\Gamma_Q$, together with the intervalley scattering rates $\Gamma_{KQ}$ and $\Gamma_{QK}$. These rate constants determine the transition propensities of the corresponding stochastic processes in the Gillespie simulation and are extracted by reproducing the experimentally measured PL decay curves. Since the $Q$ valley excitonic state has lower energy than the $K$-valley excitonic state ($\Delta E_{KQ} \approx 62$ meV), the uprising intervalley scattering rate $\Gamma_{QK}$ can be neglected by considering $k_B T << \Delta E_{KQ}$ at the low-temperature measurements, where the phonon-assisted scattering is strongly suppressed by the large energy difference. Therefore, only $\Gamma_K$, $\Gamma_Q$ and $\Gamma_{KQ}$ are included in the Monte Carlo simulation.

Using the kinetic scheme described above, $10^6$ stochastic exciton trajectories were simulated for each charge-filling condition in the simulation. The PL decay curves obtained from the Monte Carlo simulations (solid line) well reproduce the experimentally measured IX results (dots), as demonstrated in **Figure 2(d)** for each filling factor. To further elucidate the origin of the charge filling-dependent evolution of the IXs lifetime, we analyzed the kinetic parameters extracted from the simulations over the entire charge-filling range. The resulting decay rates $\Gamma_K$ (purple line) and $\Gamma_Q$ (orange line) are summarized in the upper graph of **Figure 2(e)**. Both $\Gamma_K$ and $\Gamma_Q$ exhibited similar trends as a function of filling factor. At slightly hole-filling regime $(0 < n < \sim -0.2)$, both $\Gamma_K$ and $\Gamma_Q$ decrease monotonically, indicating a gradual slowdown of IXs exciton decay. The amplitude of the non-radiative decay component in **Fig. S3(a)** increased slightly, suggesting that phonon-assisted intervalley scattering began to contribute more significantly to the decay process in the low filling regime. Upon further filling $(n > -0.2)$, both recombination rates increase rapidly and reach much higher values.

In the electron-filling regime, the recombination rates of the two components slightly increased. Meanwhile, the ratio of the initial exciton population between the $K$ and $Q$ valleys obtained from the simulation exhibits a similar trend, as presented in the lower graph of **Fig. 2(e)**. This ratio reaches the maximum during the formation of strongly correlated Mott insulating state ($n = -1$), indicating that the initial population in the $Q$ valley is more strongly suppressed than the $K$ valley. This redistribution of the exciton population suggests that the strongly correlated state suppresses the phonon-assisted nonradiative decay channel, thereby making radiative decay more dominant.

Next, we investigated the valley degree of freedom of the moiré IXs by conducting helicity-resolved PL measurements. A co-polarized pulsed laser centered at 1.72 eV was used to resonantly excite the A-exciton of $WSe_2$ in $WSe_2/WS_2$ moiré heterobilayer device. **Figure 3(a)** shows helicity-resolved PL spectra that represents the co- ($\sigma^+$) and cross-polarized ( $\sigma^-$ ) emission signal detection under $\sigma^+$ polarized excitation condition. The higher PL intensity of the $\sigma^-$-component indicates that the system exhibits negative circular polarization at the investigated filling conditions, which is consistent with the previously reported results arising from the optical selection rule in H-stacking $WSe_2/WS_2$ heterobilayer[27].

With increase of filling factor, both $\sigma^+$ and $\sigma^-$ PL intensities change with integer filling factors, as depicted in **Fig. 3(b)**. The degree of circular polarization (DOCP) defined as $\rho = \frac{I_{\sigma^+} - I_{\sigma^-}}{I_{\sigma^+} + I_{\sigma^-}}$ is calculated and plotted in **Fig. 3(c)**, where the $I_{\sigma^+}$ and $I_{\sigma^-}$ denote to the PL intensities of $\sigma^+$ and $\sigma^-$ components, respectively. The red shaded region represents the uncertainty in the DOCP calculation, which primarily originates from fluctuation in the background of PL spectra. The enhanced DOCP near $n = \pm 1$ suggests that the valley polarization is closely associated with the formation of strongly correlated states at integer fillings.

We also examined the valley relaxation time of the moiré IXs to clarify the influence of strongly correlated states on valley depolarization. The time evolution of the DOCP is described as follows.

$$\rho(t) = \frac{I_{\sigma^+}(t) - I_{\sigma^-}(t)}{I_{\sigma^+}(t) + I_{\sigma^-}(t)} \tag{2}$$

where $I_{\sigma^+}(t)$ and $I_{\sigma^-}(t)$ denote to the helicity-resolved PL decay intensities in **Fig. 4(a)**. The valley relaxation dynamics of moiré IXs monitored by $\rho(t)$ at various filling factors are shown in **Fig. 4(b)**, which can be well fitted by the double-exponential decay function:

$$\rho(t) = A_{\mathrm{r,f}}\exp\left(-\frac{t}{\tau_{\mathrm{r,f}}}\right) + A_{\mathrm{r,s}}\exp\left(-\frac{t}{\tau_{\mathrm{r,s}}}\right) \tag{3}$$

where $A_{\mathrm{r,f}}$ ($\tau_{\mathrm{r,f}}$) and $A_{\mathrm{r,s}}$ ($\tau_{\mathrm{r,s}}$) denote for the amplitudes (relaxation time) of the fast and slow components, respectively. **Figure 4(c)** summarizes the effective valley relaxation times, calculated by the equation $\tau_{\mathrm{r,eff}} = \frac{A_{\mathrm{r,f}}\tau_{\mathrm{r,f}}+A_{\mathrm{r,s}}\tau_{\mathrm{r,s}}}{A_{\mathrm{r,f}}+A_{\mathrm{r,s}}}$. As the system approaches integer fillings in both the electron and hole filling regimes, the effective valley relaxation time of the moiré exciton exhibits a modest enhancement.

It has been reported that the exciton valley relaxation in $WSe_2/WSe_2$ homobilayers is accelerated in the carrier-doped regime because the excess carriers facilitate intervalley scattering between opposite valleys[46]. In contrast, we observed a moderate increase in the valley relaxation time near the integer filling factors in the $WSe_2/WS_2$ heterobilayer. To rationalize the biexponential valley dynamics, we consider a minimal bright–dark reservoir picture, as schematically illustrated in **Fig. 4(d)**. The first is the valley relaxation pathway associated with intervalley coupling between the $|+K\rangle$ and $|-K\rangle$ valley pseudospins of bright moiré IX. Because these configurations are time-reversal partners and are nearly degenerate at zero magnetic field, they can be coupled by the long-range electron-hole ($e$-$h$) exchange interaction[47], providing a fast depolarization process. The second pathway corresponds to a slower depolarization process involving spin-flip and phonon-assisted scattering from the bright $K$ excitonic state to the dark $Q$ excitonic state[48]. Because this transition involves a large momentum mismatch, phonon assistance is required to satisfy momentum conservation, resulting in a substantially slower relaxation process. Following this transfer, the exciton population accumulated in the dark $Q$ excitonic state can be viewed as a long-lived excitonic reservoir[49]. Owing to its reduced oscillator strength, valley polarization stored in this dark-state reservoir can persist over a substantially longer timescale. Such reservoir-mediated dynamics provide a microscopic origin for the slow depolarization component observed on the tens-of-nanoseconds timescale in the experiments.

Near integer fillings, the formation of correlated insulating states in the moiré flat minibands might localize the IXs and renormalize their coherent spectral weight and dielectric environment. These effects might reduce the effective $e$-$h$ exchange interaction, thereby suppressing the fast depolarization channel. Monte Carlo analysis of the IX population dynamics further reveals a moderate reduction in the effective phonon-assisted nonradiative process near $n = \pm 1$ (**Fig. S3**), which is consistent with a weakened exciton–phonon coupling in the correlated regime. The results illustrate the suppressed bright to dark IXs conversion and a correspondingly prolonged slow depolarization process. Similar behavior is observed in device H2 (**Fig. S4**), including an enhancement of the valley relaxation time near $n = -1$.

## Conclusion

In summary, we investigated the charge filling-dependent valley relaxation dynamics of IXs in $WSe_2/WS_2$ moiré heterobilayer system. We revealed anomalous behaviors of IXs recombination dynamics at integer filling of moiré site $n = \pm 1$, accompanied by enhanced excitonic coherence. In low charge-filling regime, increasing the carrier density slows the exciton recombination rates and decreases the decoherence through enhanced intervalley scattering. However, the emergence of the Mott insulating state at $n = \pm 1$ alters this trend because of carrier localization and charge incompressibility. Meanwhile, the valley relaxation time of moiré IXs is prolonged because of the suppression of scattering-mediated pseudospin flipping and reduced $e$-$h$ exchange interaction. These findings establish strongly correlated states as a powerful knob for modulating the valley degree of freedom of moiré IXs, and they position the $WSe_2/WS_2$ moiré heterobilayer as a promising platform for valleytronic applications, including valley-based quantum memory and quantum computation.

## Methods

**Sample fabrication:** Monolayers of semiconductors, $WSe_2$, and $WS_2$ were prepared using a mechanical exfoliation method from bulk crystals. Few-layer graphene (as electrodes) and $h$-BN layers were prepared by mechanical exfoliation. A $WSe_2/WS_2$ heterobilayer device with three graphene electrodes encapsulated by $h$-BNs was fabricated using an all-dry-transfer technique. A polydimethylsiloxane (PDMS) stamp and polyvinyl chloride (PVC) film were used during the transfer[11]. The fabricated device was thermally annealed at 225 °C for 1 h in an Ar/$H_2$ (97%/3%) atmosphere to reduce interfacial contamination, improve the contact between the monolayers, and remove residual PVC from the device surface.

**Steady-state PL measurements:** A linearly polarized continuous-wave (cw) semiconductor laser with a photon energy of 2.33 eV was used as the excitation source for the PL measurements of the intralayer excitons (IXs). The laser was focused on the device surface using a ×50 objective lens, resulting in a spot size of approximately 1.5 μm. The PL spectra of the $WSe_2/WS_2$ heterobilayer device were collected using a cooled charge-coupled device (CCD) coupled to a spectrometer with a spectral resolution of 0.5 meV. The devices were cooled from 298 to 5.8 K using a cryogen-free cryostat.

**Time-resolved PL measurements:** A pulsed supercontinuum light source with a photon energy of 1.72 eV (pulse width of ~100 ps and repetition rate of 4 MHz) was used to perform time-resolved PL measurements. The combination of long- and short-pass filters with cut-off energies of 1.38 and 1.58 eV, respectively, was used to obtain the PL decay signals from the $WSe_2/WS_2$ IXs.

**Supporting Information**

The supporting Information is available free of charge at: Analytical procedures of the discrete Monte Carlo simulation; Doping-dependent $\sigma^+$ PL spectra of IXs and intralayer exciton PL spectra of $WSe_2$ in device H2; amplitudes of the fast ($A_f$) and slow ($A_s$) components extracted from the biexponential decay function using eq. (1), and the corresponding decay lifetimes ($\tau_f$ and $\tau_s$) in device H1; Valley relaxation time at different filling factors in device H2.

**Author Information**

**Corresponding Author**

**Kazunari Matsuda** - *Institute of Advanced Energy, Kyoto University, Uji, Kyoto 611-0011, Japan*; Email: matsuda@iae.kyoto-u.ac.jp

**Authors**

**Zhiwei Li** - *Institute of Advanced Energy, Kyoto University, Uji, Kyoto 611-0011, Japan*

**Haonan Wang** - *Institute of Advanced Energy, Kyoto University, Uji, Kyoto 611-0011, Japan*

**Yubei Xiang** - *Institute of Advanced Energy, Kyoto University, Uji, Kyoto 611-0011, Japan*

**Wataru Idehara** - *Institute of Advanced Energy, Kyoto University, Uji, Kyoto 611-0011, Japan*

**Hailin Liu** - *Institute of Advanced Energy, Kyoto University, Uji, Kyoto 611-0011, Japan*

**Shinichiro Matano** - *Institute of Advanced Energy, Kyoto University, Uji, Kyoto 611-*

*0011, Japan*

**Kenji Watanabe** - *Research Center for Electronic and Optical Materials, National Institute for Materials Science, 1-1 Namiki, Tsukuba 305-0044, Japan*

**Takashi Taniguchi** - *Research Center for Materials Nanoarchitectonics, National Institute for Materials Science, 1-1 Namiki, Tsukuba 305-0044, Japan*

The authors declare no competing financial conflicts of interest.

**Author Contributions**

Z.L., W.I., K.W., and T.T. contributed to the fabrication of the devices studied here. Z.L., H.W., and K.M. designed the experiments, which were performed by Z.L. and H.W. Data analysis was performed by Z.L., H.W., Y. X., and K.M. The draft was written by Z.L., H.W., Y. X.S.M. and K.M. All authors contributed to the review and editing of the manuscript. The project was supervised by K. M.

**Data availability**

The data presented in this paper and the supplementary materials are available from the corresponding author upon reasonable request.

**Acknowledgements**

This work was supported by the JSPS KAKENHI (Grant Numbers JP20H05664, JP21H05232, JP21H05235, JP22K18986, JP23H05469, JP23KJ1381, JP25K23547,

JP25H00417, JP25K24588, JP25K24710, JP26K17758, and JP26K21849), and the JST CREST Program (Grant Number JPMJCR24A5). K.W. and T.T. acknowledge support from the World Premier International Research Center Initiative (WPI), MEXT, Japan.

**Competing interests**

The authors declare that they have no competing interests.

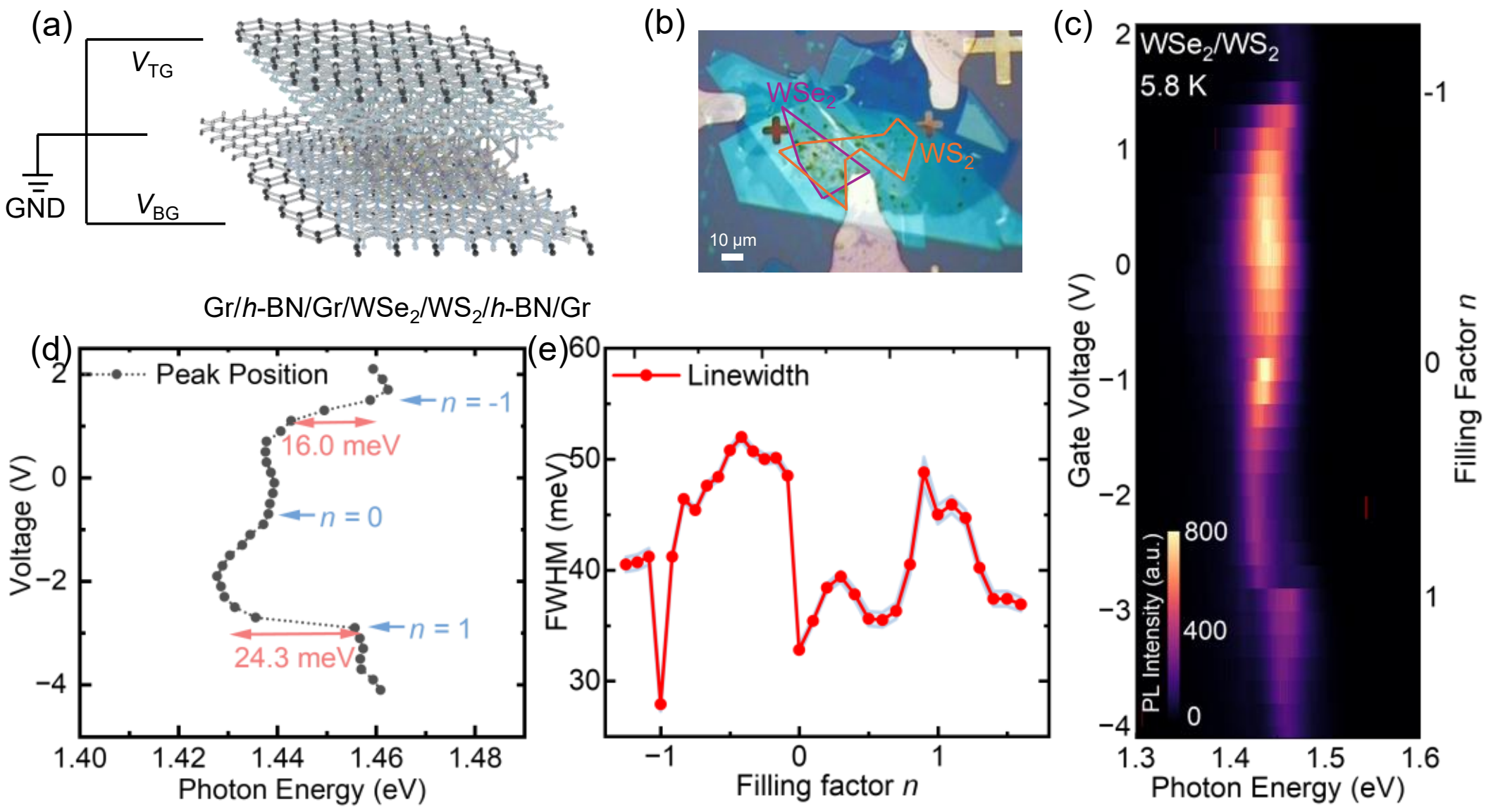

**Figure 1.** (a) Schematic illustration of the dual-gated $WSe_2/WS_2$ heterostructure device. The top and back gates are denoted by TG and BG, respectively. (b) Optical image of the fabricated device, with the scale bar of 10 μm in the image. The outlines in the image indicate the monolayer regions of $WSe_2$ (purple) and $WS_2$ (orange). (c) Charge filling dependent PL spectra of IXs measured under CW laser excitation with a photon energy of 2.33 eV. (d) Peak positions of IXs emission as a function of charge filling, extracted from the Voigt function fitting of the corresponding PL spectra. (e) Full width at half maximum (FWHM) of IXs emission peak extracted from the Voigt function fitting procedures.

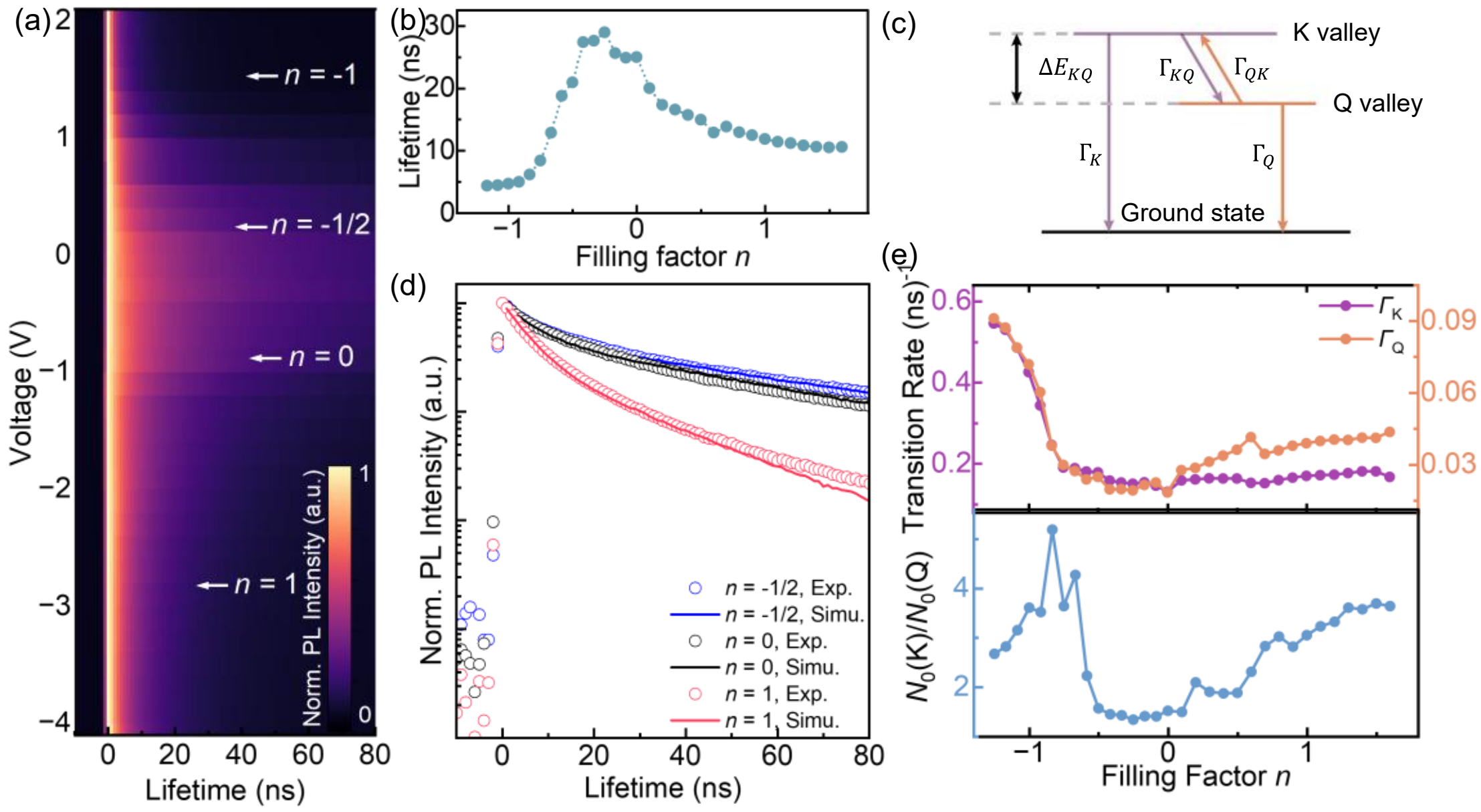

**Figure 2.** (a) Contour plot of time-resolved PL decay curves at various filling factors, with the excitation energy of the pulsed laser centered at 1.72 eV. (b) Extracted effective lifetime of the IXs as a function of filling factor. (c) Schematic illustration of the three-level system with $K$ and $Q$ valleys adopted in the Monte Carlo simulation for IXs dynamics. (d) Time-resolved PL decay curves of IXs at the filling factors of $n = 0,\ 1$ and $-1/2$, indicated by the dots. The calculated curves by the Monte Carlo simulation are also shown, as indicated by the lines. (e) Upper panel: transition rates $\Gamma_K$ (purple), and $\Gamma_Q$ (orange) of IXs from $K/Q$ valley to ground states. Lower panel: ratio of initial IXs populations in $K/Q$ valley as the function of charge filling.

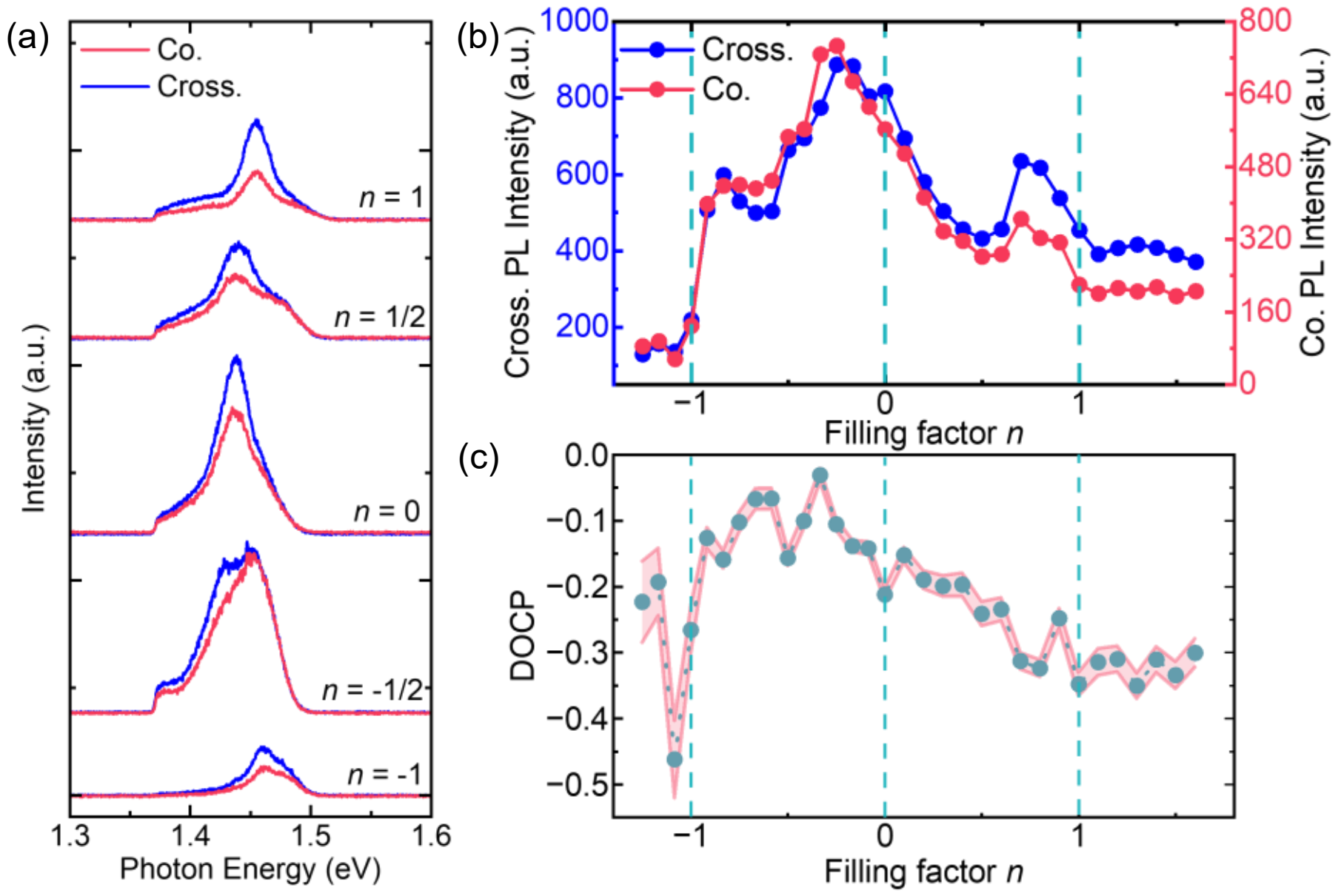

**Figure 3.** (a) Polarization-resolved PL spectra of $WSe_2/WS_2$ IXs at various filling factors under $\sigma^+$ excitation. Noted that the sharp spectral edges around 1.36 eV originate from the cutoff wavelengths of the short-pass filters and are therefore unrelated to intrinsic IX emission features. (b) Charge filling dependence of PL intensity of $\sigma^+$ and $\sigma^-$ components in the spectra. (c) DOCP of the IXs emission as a function of charge filling, calculated from the $\sigma^+$ and $\sigma^-$ PL intensities. The filling-dependent DOCP reflects the modulation of valley polarization and intervalley relaxation processes by the correlated electronic states in the $WSe_2/WS_2$ moiré heterostructure.

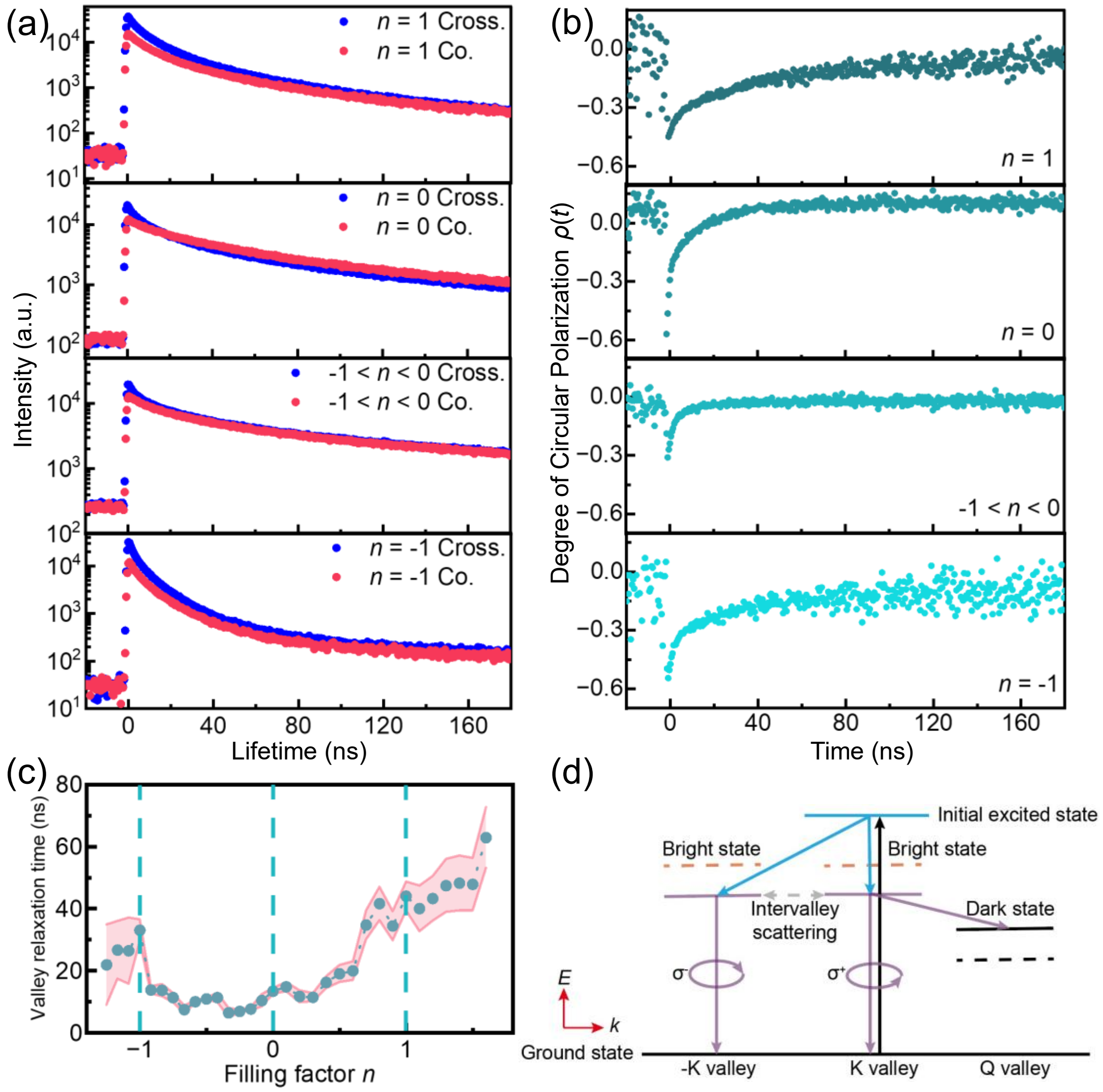

**Figure 4.** (a) Helicity-resolved PL decay curves of IXs measured under $\sigma^+$ excitation at representative charge-filling conditions. The $\sigma^+$ and $\sigma^-$ decay components reveal the filling-dependent evolution of the valley polarization. (b) Evolution of the degree of circular polarization (DOP) $\rho(t)$ under different charge filling conditions, calculated from the helicity-resolved PL intensities. The DOCP decay reflects the loss of the initial valley polarization through intervalley relaxation and pseudospin-flipping process. (c) Valley relaxation times extracted from $\rho(t)$ as a function of filling factor. The pronounced increase in the relaxation time near the integer filling indicates a suppression of valley depolarization and a prolonged preservation of the IX valley polarization. (d) Schematic illustration of the proposed mechanism responsible for the enhanced valley relaxation of IXs.

# Supporting Information for

# Charge Filling-Dependent Valley Polarization and its Dynamics Driven by Correlated Inter-layer Exciton States in $WSe_2/WS_2$ Heterobilayer

*Zhiwei Li[1], Haonan Wang[1], Yubei Xiang[1], Wataru Idehara[1], Hailin Liu[1], Shinichiro Matano[1], Kenji Watanabe[2], Takashi Taniguchi[3], and Kazunari Matsuda[1]**

[1]Institute of Advanced Energy, Kyoto University, Uji, Kyoto 611-0011, Japan

[2]Research Center for Electronic and Optical Materials, National Institute for Materials Science, 1-1 Namiki, Tsukuba, Ibaraki 305-0044, Japan

[3]Research Center for Materials Nanoarchitectonics, National Institute for Materials Science, 1-1 Namiki, Tsukuba, Ibaraki 305-0044, Japan

Corresponding author: Kazunari Matsuda

E-mail: matsuda@iae.kyoto-u.ac.jp

**Supplementary Note 1**

In **Fig. S2(a)**, blueshifts of 16 and 9.7 meV were observed at the $n = -1$ and $-2$ in $\sigma^+$ PL spectra. The direction of charge carrier filling was independently identified using the change of PL spectral peaks arising from intralayer $WSe_2$ excitonic features (**Fig. S2(b)**), which is consistent with the previous study[50]. In monolayer $WSe_2$, free exciton peak appears at the highest emission energy, followed by positively and negatively charged excitons (trions) at the lower energies. Specifically, the emission energy of the positive trion is approximately 1.685 eV, whereas that of the negative trion is approximately 1.675 eV[51].

**Supplementary Note 2**

The excitation power of the pulse laser used in the helicity and time-resolved measurements was approximately 120 nW, which was far below the saturated power of the $WSe_2/WS_2$ IXs emission[27]. Under this weak-excitation condition, the number of photoexcited IXs remains below the maximum occupancy available at the moiré sites. Thus, the effect of high exciton density associated with site saturation and/or state filling are neglected in the simulation.

The stochastic time evolution for the simulation is implemented using the Gillespie algorithm. Each simulated trajectory starts at $t = 0$ and is propagated over a total temporal window of 80 ns, which corresponds to the time range of the experimental time-resolved PL measurements. No fixed simulation time step is introduced. Instead, the waiting time between two successive stochastic events is randomly sampled

according to:

$$\Delta t = -\frac{\ln r}{\Gamma_{\mathrm{tot}}},$$

where $r$ is a uniformly distributed random number between 0 and 1 and $\Gamma_{\mathrm{tot}}$ represents the sum of the transition rates accessible from the currently occupied excitonic state. The stochastic trajectories are subsequently accumulated along the experimental time axis to simulate the PL decay profiles.

In addition, the scattering constant $\gamma_0$ is allowed to vary with charge carrier filling. This treatment is motivated by the pronounced charge filing dependence of the extracted $\Gamma_Q$, suggesting that the relaxation kinetics associated with the $Q$-valley excitonic state are strongly modified by the filling factor.

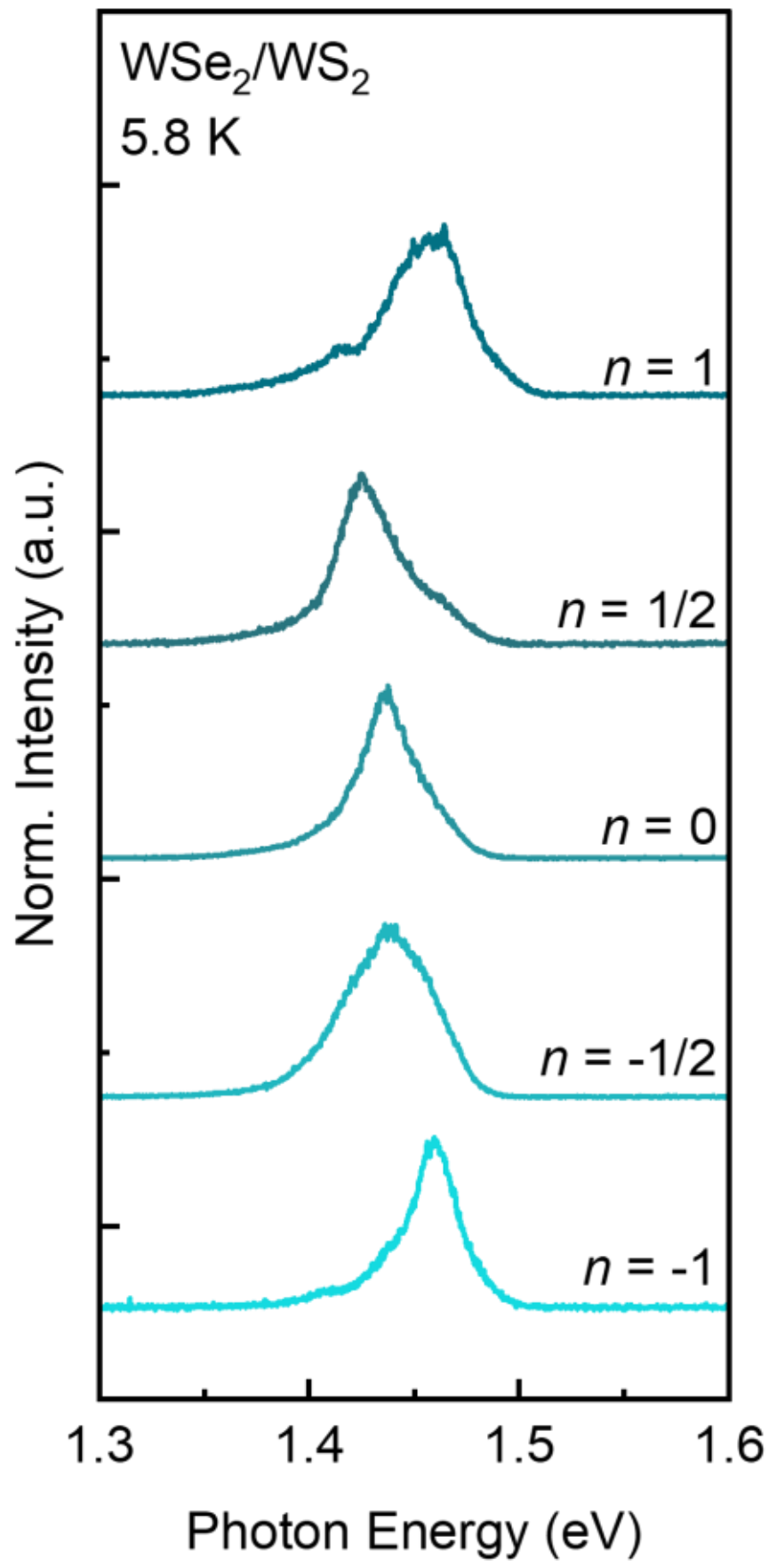

**Supplementary Figure 1.** Normalized PL spectra of IXs in the H1 $WSe_2/WS_2$ heterostructure device measured at various charge filling factors.

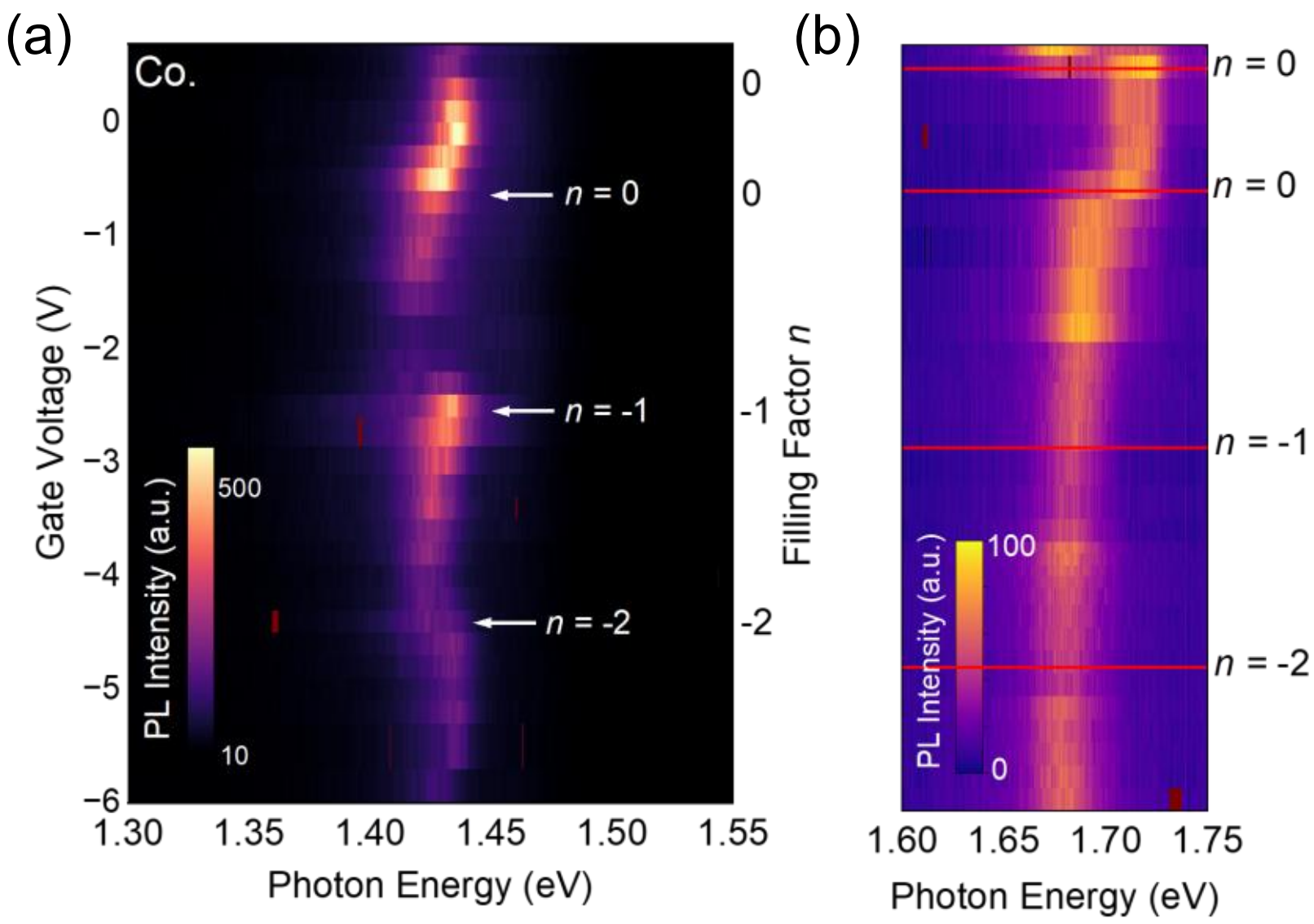

**Supplementary Figure 2**. (a) Charge filling dependent PL spectra of IXs in device H2. (b) Charge filling dependent $WSe_2$ intralayer exciton PL spectra adopted to precisely determine the charge filling.

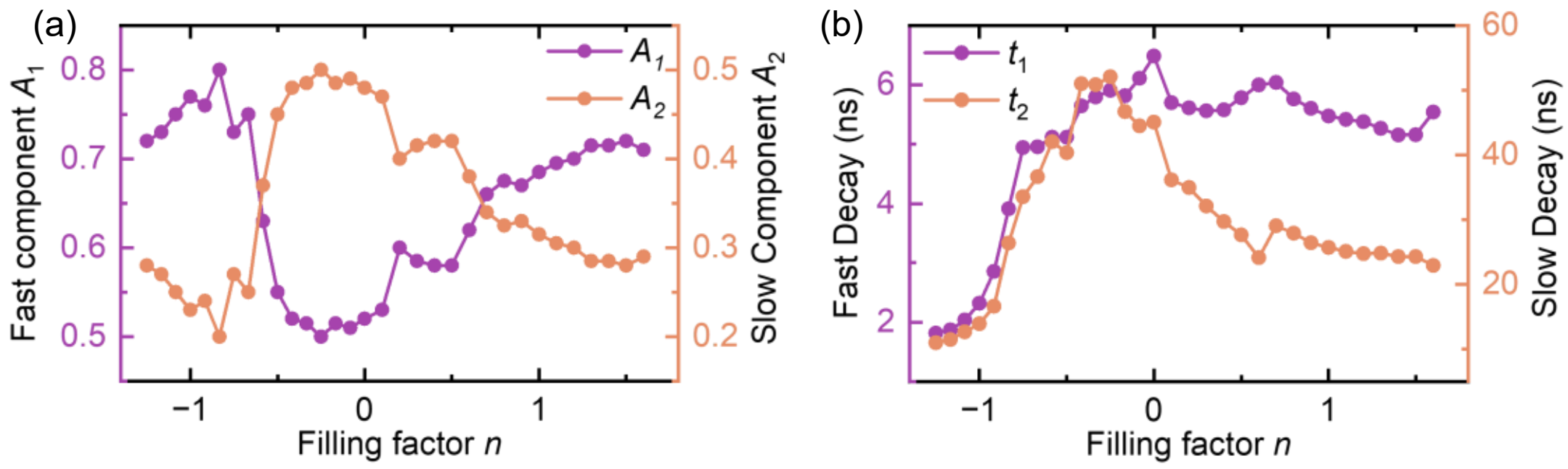

**Supplementary Figure 3**. (a) Amplitudes of the fast component ($A_f$) and slow decay component ($A_s$) extracted from the double exponential decay using eq. (1). (b) Corresponding decay times of the fast component ($\tau_f$) and slow component ($\tau_s$).

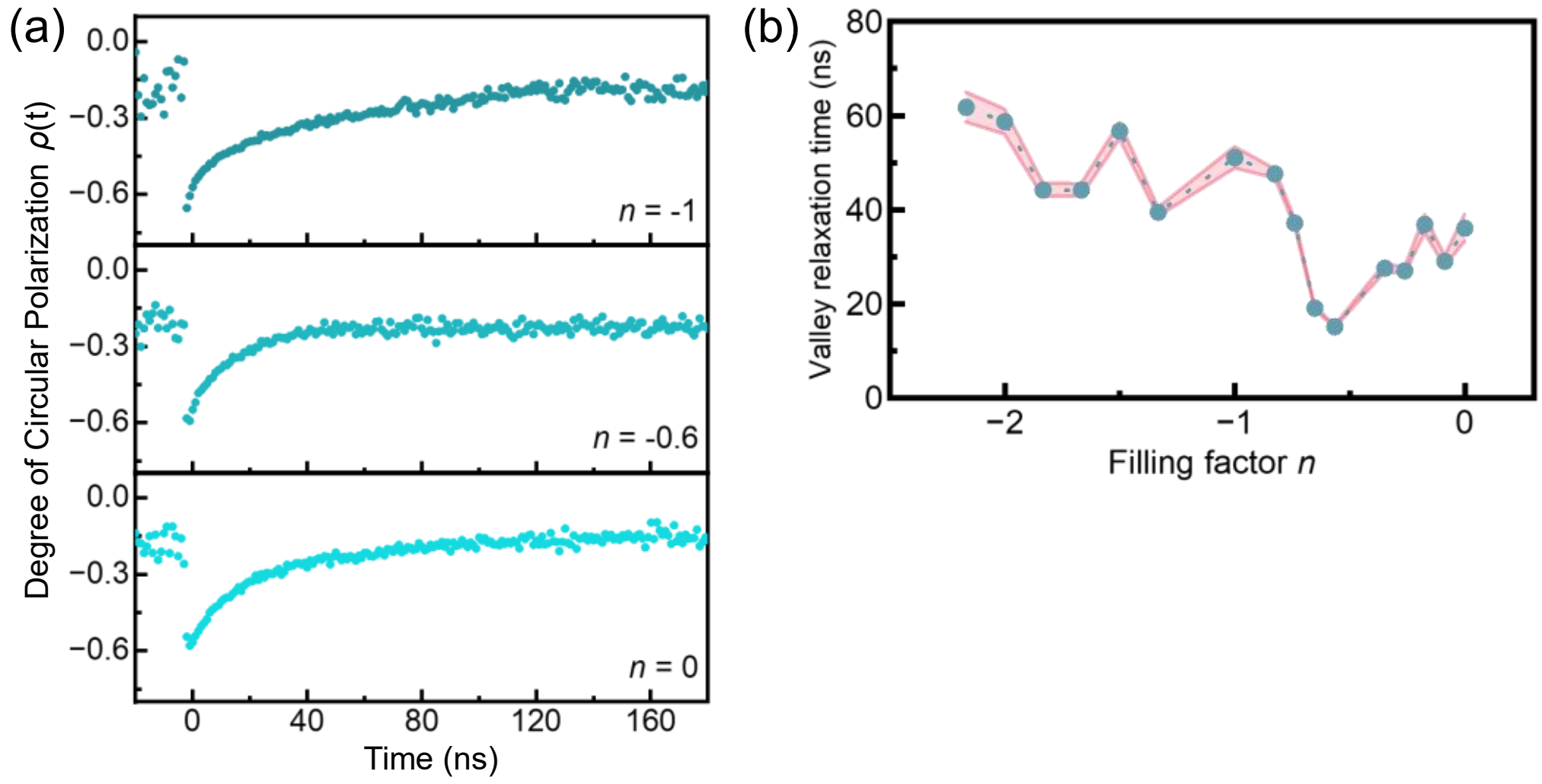

**Supplementary Figure 4**. (a) Evolution of the degree of circular polarization (DOCP) $\rho(t)$ under different charge filling conditions for device H2. (b) Extracted valley relaxation time as a function of filling factor, obtained by fitting the corresponding DOCP decay curves.